\documentclass[aps,prl,twocolumn]{revtex4-1}
\usepackage[utf8]{inputenc}
\usepackage[T1]{fontenc}
\usepackage[english]{babel}
\usepackage{amssymb,amsmath,mathtools} % Math
\usepackage{graphicx} 
\usepackage{xcolor}
\usepackage{bbold}
\usepackage{braket}
\usepackage[headings]{fullpage}
\usepackage{hyperref}
\usepackage{slashed}
\usepackage[titletoc,toc,title]{appendix}
\usepackage[hang,small,bf]{caption}
\usepackage[nottoc,numbib]{tocbibind}
\usepackage[export]{adjustbox}
\usepackage{xcolor}
\usepackage{array}
\usepackage{booktabs}
\usepackage{hyperref}
\usepackage{amsthm}
\usepackage{lineno}

\begin{document}
	\newcommand{\be}{\begin{eqnarray}}
		\newcommand{\ee}{\end{eqnarray}}
	\newcommand{\del}{\partial}
	\newcommand{\nn}{\nonumber}
	\newcommand{\STr}{{\rm Str}}
	\newcommand{\Sdet}{{\rm Sdet}}
	\newcommand{\Pf}{{\rm Pf}}
	\newcommand{\mat}{\left ( \begin{array}{cc}}
		\newcommand{\emat}{\end{array} \right )}
	\newcommand{\vect}{\left ( \begin{array}{c}}
		\newcommand{\evect}{\end{array} \right )}
	\newcommand{\tr}{{\rm Tr}}
	\newcommand{\hm}{\hat m}
	\newcommand{\ha}{\hat a}
	\newcommand{\hz}{\hat z}
	\newcommand{\hze}{\hat \zeta}
	\newcommand{\hx}{\hat x}
	\newcommand{\hy}{\hat y}
	\newcommand{\tm}{\tilde{m}}
	\newcommand{\ta}{\tilde{a}}
	\newcommand{\U}{\rm U}
	\newcommand{\D}{\slashed{D}}
	\newcommand{\hc}{^\dagger}
	\newcommand{\inv}{^{-1}}
	\newcommand{\diag}{{\rm diag}}
	\newcommand{\sign}{{\rm sign}}
	\newcommand{\ct}{\tilde{c}}
	\newtheorem{theorem}{Conjecture}
	\newcommand{\eins}{\leavevmode\hbox{\small1\kern-3.8pt\normalsize1}}
	
	\title{Fundamental Lack of Information in Observed Disease and Hospitalization Data}

	\author{Adam Mielke}\email{admi@dtu.dk}\affiliation{Dynamical Systems, Technical University of Denmark, Asmussens Allé, 303B, 2800 Kgs.\ Lyngby, Denmark}
	\author{Lasse Engbo Christiansen}\email{lsec@ssi.dk}\affiliation{Statens Serum Institut, Artillerivej 5, 2300 Copenhagen S, Denmark}
	
	\date{\today}
	
	\begin{abstract}
		We present a proof based on SEIR-models that shows it is impossible to identify the level of under reporting based on traditional observables of the disease dynamics alone. This means that the true attack rate must be determined through other means.
	\end{abstract}
	
	\date{\today}
	\maketitle
	
	\section{Introduction}
	Lack of information is a central problem in human epidemiology that has long been studied \cite{UnderReporting1, UnderReporting2, UnderReporting3, UnderReporting4, UnderReporting5, UnderReporting6, UnderReporting7, UnderReporting8}. The number of sick people is filtered through the amount that are being tested, which also affects the perceived risk of admission to the hospital. Considering the role of observables is therefore key to understanding how the true number of infected behaves.
	
	Because of the non-linear nature of disease spread, it is tempting to think that it is not possible to scale out a factor in front of the number of infected and that a constant level of under reporting would be detectable simply through the dynamics of the disease over time. We here show that this is not the case. Even a constant level of under reporting can be hidden from the observables, which requires the true prevalence to be found in other ways, such as blood donor seroprevalence \cite{SSIdonors} or a large unbiased random sample of the population \cite{Sweden, UKSample}.
	
	We work in an SEIR-model framework. SIR-type models are well-established and widely used in disease-modelling \cite{SIR1, SIR2, SIR3, SpatialEpidemicsPoland, SpatialEpidemicsStatMech, EkspertgruppenRapporter, SpatialEpidemicsGraph, SpatialEpidemicsFluid1, SpatialEpidemicsFluid2, SpatialEpidemicsFluid3, Liu, SIR4, SIR5}.
	It is a simple ODE-setup that allows many explicit calculations, but still provides key insights about disease spread.
	
	\section{Observables}
	Let us consider the following setup where susceptible ($S$), exposed ($E$), infectious ($I$), hospitalized ($H$), and recovered ($R$) are all vectors representing the distribution of the population on age groups and states. We assume that there is fundamental age structured SEIR-model with a hospital state representing the actual disease progression
	\begin{align}\label{Eq:SEIR-model}
		\begin{split}
			\dot{S} &= - \diag(S)\beta I\\
			\dot{E} &=   \diag(S) \beta I - \eta E\\
			\dot{I} &= \eta E - \gamma I\\
			\dot{H} &= \gamma \diag(p_H) I - \diag(\sigma) H\\
			\dot{R} &= (\eins - \diag(p_H))\gamma I + \sigma H
		\end{split}\ .
	\end{align}
	Here $\eta$ and $\gamma$ are scalar parameters, whereas $p_H$ and $\sigma$ are vectors, reflecting different hospitalization risks and duration of admissions for different age groups. The contact matrix $\beta$ is symmetric and non-negative in all elements (at least in general). All the parameters and elements of matrices are non-negative.
	We use the normalization
	\begin{eqnarray}\label{Eq:Norm}
		S + E + I + H + R&=& \nu\ ,
	\end{eqnarray}
	where $\nu$ is the population distribution, and $\sum_j\nu_j=1$.
	
	Derived from these states and parameters are the following observables
	\begin{align}\label{Eq:SEIR-model:Observables}
		\begin{split}
			N_{\rm pos} &= \eta \diag(a) E\\
			N_{\rm adm} &= \gamma \diag(p_H) I\\
			N_{\rm hosp} &= H
		\end{split}
	\end{align}
	which determine the information we have about the system. $N_{\rm pos}$ is the number of new test positive, and $a$ is a vector representing under reporting for each age group. In this way we only have partial information about the disease dynamics. Unless there is an extraordinary number of false positives, we have $a_j\leq 1$. $N_{\rm adm}$ and $N_{\rm hosp}$ are the number of new admissions and people in hospital, respectively. We assume that we have perfect information about the hospital, but because of the under reporting, there will be an observed risk of admission
	\begin{eqnarray}
		p_{H, Obs} &=& \diag(p_H) \diag(a)\inv\ .
	\end{eqnarray}
	We assume that $\eta$ and $\gamma$ may be determined with a high degree of accuracy and precision by studying patients, from interviews of known cases and emerging literature.
	
	The goal of this paper is to show that the above limitations on the observables mean that the dynamics cannot be uniquely determined without other sources of information.
	
	\section{Degrees of Freedom}
	Let us investigate which transformations we can make that keeps the dynamics invariant. It turns out that all equations but the one for $R$ remain invariant under the transformation
	\begin{align}\label{Eq:SEIR-model:Trans}
		\begin{split}
			S &= g \tilde{S}\\
			E &= g \tilde{E}\\
			I &= g \tilde{I}\\
			p_H &= \tilde{p}_H g \inv\\
			\beta &= \tilde{\beta} g\inv
		\end{split}\ .
	\end{align}
	Mathematically, $g$ can be a general matrix, but in most realistic systems it will be diagonal. For now we keep the transformation general, but it will be related to under reporting in the end. Note that we do not rescale $H$, because it is known exactly.
	The transformation must preserve $0\leq \tilde{p}_H \leq 1$ for all elements of $\tilde{p}_H$.
	We find
	\begin{align}\label{Eq:SEIR-model:Reduced}
		\begin{split}
			g\dot{\tilde{S}} &= - g\diag(\tilde{S})\tilde{\beta}g\inv g \tilde{I}\\
			g\dot{\tilde{E}} &= g\diag(\tilde{S}) \beta g\inv g \tilde{I} - \eta g \tilde{E}\\
			g\dot{\tilde{I}} &= \eta g \tilde{E} - \gamma g \tilde{I}\\
			\dot{H} &= \gamma \diag(\tilde{p}_H) \tilde{I} - \sigma H\\
			\dot{R} &= (g-\diag(p_H))\gamma \tilde{I} + \sigma H
		\end{split}\ .
	\end{align}
	Note that $\beta$ and $g$ do not commute, which is why we also have to transform $\beta$ as well. The above can be reduced to
	\begin{align}\label{Eq:SEIR-model:Reduced2}
		\begin{split}
			\dot{\tilde{S}} &= - \diag(\tilde{S})\tilde{\beta} \tilde{I}\\
			\dot{\tilde{E}} &=   \diag(\tilde{S}) \tilde{\beta} \tilde{I} - \eta \tilde{E}\\
			\dot{\tilde{I}} &= \eta \tilde{E} - \gamma \tilde{I}\\
			\dot{H} &= \gamma \diag(\tilde{p}_H) \tilde{I} - \sigma H\\
			\dot{R} &= (g-\diag(p_H))\gamma \tilde{I} + \sigma H
		\end{split}\ .
	\end{align}
	If we set $g = \diag(a)\inv$, where the inverse is just because of the definition in Equation \eqref{Eq:SEIR-model:Trans}, we find
	\begin{align}\label{Eq:SEIR-model:ObservablesRescale}
		\begin{split}
			N_{\rm pos} &= \eta \tilde{E}\\
			N_{\rm adm} &= \gamma \diag(\tilde{p}_H) \tilde{I}\\
			N_{\rm hosp} &= H
		\end{split}
	\end{align}
	This means that it is impossible to distinguish between the models \eqref{Eq:SEIR-model} and \eqref{Eq:SEIR-model:Reduced2}  based on the observables. Note that we also need
	\begin{eqnarray}\label{Eq:NormRescale}
		\tilde{S} + \tilde{E} + \tilde{I} + H &\leq& \nu\  .
	\end{eqnarray}
	Otherwise \eqref{Eq:SEIR-model:Reduced} is easily identifiable as a wrong representation. This sets a lower bounds on the elements of $a$ that is greater than zero. The bound also becomes tighter if a large fraction of the population is tested.
	
	\subsection{Simulation}
\begin{figure*}
		\centering
		\includegraphics[width=0.9\linewidth]{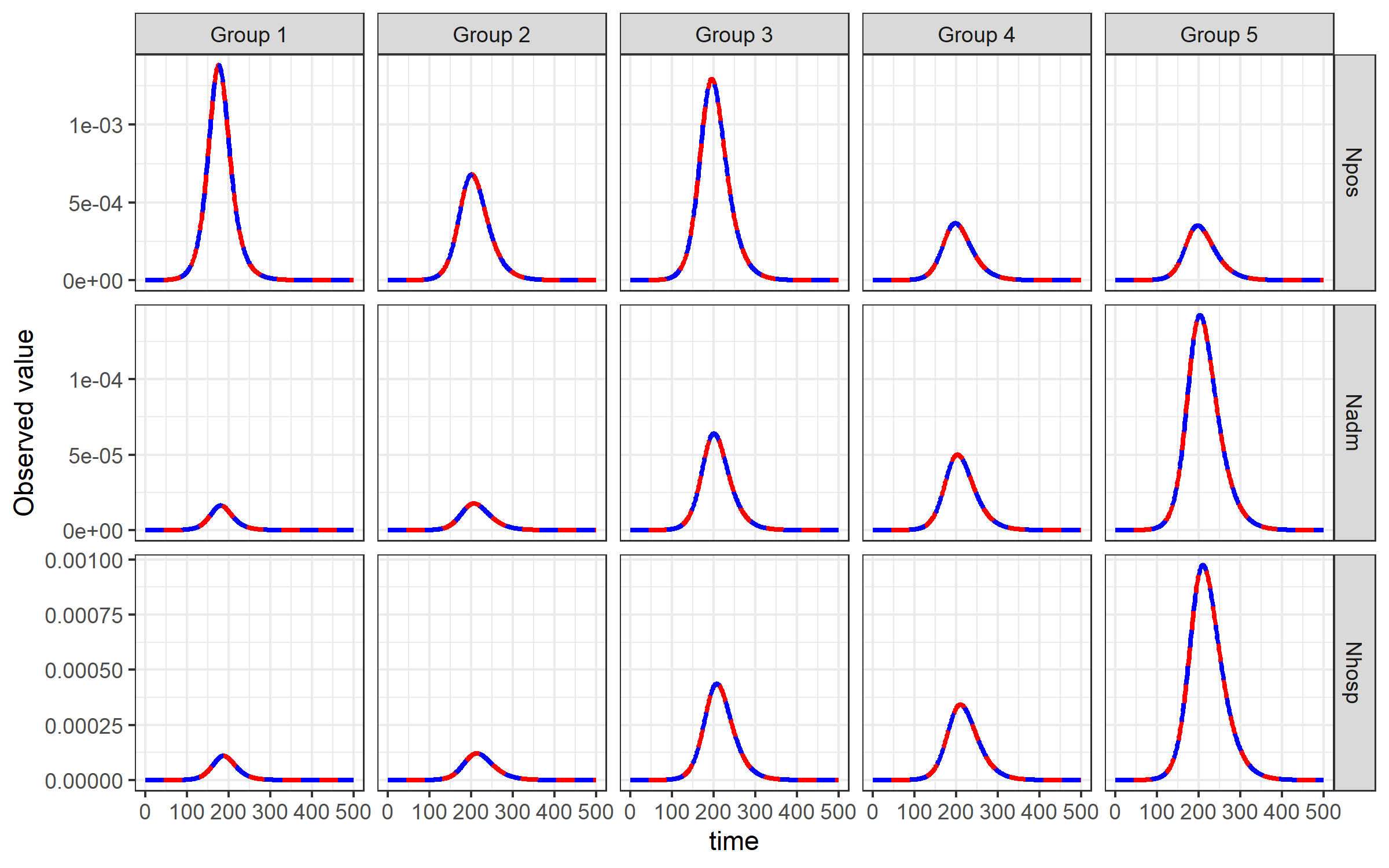}
		\caption{SEIR-model with 5 groups. Plotted are the Models 1 and 2 with their respective observables as described in the Simulation section.
			The fact that these coincide illustrates our point that partial information from positive tests is indistinguishable from a system with full information and other parameters and initial conditions.}
		\label{Fig:SEIR}
	\end{figure*}

	To illustrate this, we run two simulations:
	\begin{itemize}
		\item Model 1: Observing a fraction of the true positive
		\begin{itemize}
			\item Initial conditions
			\begin{align*}
				S(0) = \nu-\epsilon\ ,\ E(0) = \epsilon\ ,\ I(0) = 0\ ,\ R(0) = 0
			\end{align*}
			\item Parameters $a$, $\beta$, $\eta$, $\gamma$, and $p_H$, see Equation \eqref{Eq:ParameterValues}.
			\item Observables
			\begin{align*}
				\begin{split}
					N_{\rm pos} &= \eta \diag(a) E\\
					N_{\rm adm} &= \gamma \diag(p_H) I\\
					N_{\rm hosp} &= H
				\end{split}
			\end{align*}
		\end{itemize}
		\item Model 2: Rescaled so that it is assumed that all true positive are observed
		\begin{itemize}
			\item Initial conditions
			\begin{align*}
				S(0) = \diag(a)(\nu-\epsilon)\ &,\ E(0) = \diag(a)\epsilon\ ,\\
				I(0) = 0\ &,\ R(0) = 0
			\end{align*}
			\item Parameters $\tilde{\beta} = \beta \diag(a)\inv$, $\eta$, $\gamma$, and $\tilde{p}_H = p_H\diag(a)\inv$, again see Equation \eqref{Eq:ParameterValues}.
			\item Observables
			\begin{align*}
				\begin{split}
					N_{\rm pos} &= \eta E\\
					N_{\rm adm} &= \gamma \tilde{p}_H I\\
					N_{\rm hosp} &= H
				\end{split}
			\end{align*}
		\end{itemize}
	\end{itemize}
	We choose generated parameters that are similar to real-world data
	\begin{align}\label{Eq:ParameterValues}
		\begin{split}
			\beta &= \left(\begin{matrix}
				1.261 & 0.059 & 0.094 & 0.070 & 0.051\\
				0.059 & 1.020 & 0.052 & 0.083 & 0.031\\
				0.094 & 0.052 & 0.906 & 0.076 & 0.008\\
				0.070 & 0.083 & 0.076 & 0.876 & 0.030\\
				0.051 & 0.031 & 0.008 & 0.030 & 0.641
			\end{matrix}\right)\\
			\nu &= \left(\begin{matrix}
				0.243 & 0.198 & 0.228 & 0.126 & 0.205
			\end{matrix}\right)^T\\
			p_H &= \left(\begin{matrix}
				0.0067 & 0.0183 & 0.0498 & 0.1353 & 0.3679
			\end{matrix}\right)^T\\
			\epsilon &= \left(\begin{matrix}
				2.425\cdot 10^{-6} & 0 & 0 & 0 & 0
			\end{matrix}\right)^T\\
			a &= \left(\begin{matrix}
				0.556 & 0.691 & 0.990 & 0.979 & 0.899
			\end{matrix}\right)^T\\
			& \eta = 1/4.3\ ,\ \gamma = 1/5.3\ ,\ \sigma = 1/7
		\end{split}
	\end{align}
	See Figure \ref{Fig:SEIR} for the comparisons of the two models.

	\subsection{Remark on the symmetry of $\beta$}
	The transformation $\beta = \tilde{\beta} g\inv$ makes $\tilde{\beta}$ asymmetric, but the alternative transformation
	\begin{align}\label{Eq:SEIR-model:Trans2}
		\begin{split}
			\diag(S) &= g\diag(\hat{S})g\\
			E &= g \tilde{E}\\
			I &= g \tilde{I}\\
			p_H &= \tilde{p}_H g \inv\\
			\beta &= g\inv \hat{\beta} g\inv
		\end{split}\ ,
	\end{align}
	which leads to
	\begin{align}\label{Eq:SEIR-model:Reduced3}
		\begin{split}
			\diag(\dot{\hat{S}})g &= - g \diag \left(\diag(\hat{S})\tilde{\beta} \tilde{I}\right)\\
			\dot{\tilde{E}} &= \diag(\tilde{S}) \tilde{\beta} \tilde{I} - \eta \tilde{E}\\
			\dot{\tilde{I}} &= \eta \tilde{E} - \gamma \tilde{I}\\
			\dot{H} &= \gamma \tilde{p}_H \tilde{I} - \sigma H\\
			\dot{R} &= (g-p_H)\gamma \tilde{I} + \sigma H
		\end{split}\ ,
	\end{align}
	does have symmetric $\beta$.
	The observables remain the same, and for diagonal $g$, the equation for $S$ is independent of $g$, and the observables are thus invariant under the transformation.
	
	\subsection{Note on the generality of the result}
	In this paper we show the calculation explicitly for an SEIR-model, but we expect the result to hold for a much larger class of models. As long as the parameters in front of each state are distinct, each non-linear term has a parameter that is not directly observable, and an individual can maximally occupy a given state once, similar transformations are possible. A term like $\diag(S)\beta I^\alpha$ poses for instance no problem, but $\gamma I^\alpha$ on its own does, as there nowhere to absorb the transformation $a$ when the non-linearity is in the delay term.
	
	Additionally, agent-based or network models have the same degrees of freedom in the probabilities of infection. It is simply important to take into account the level of stratification that the model works on. E.g.\ if a test can be attributed to a specific municipality, then $a$ must contain each municipality as well. However, the bound on the parts of $a$ that are tested less becomes worse, so a greater understanding is not obtained. (The extreme case is knowledge on an individual basis, which has the bound $a=1$ for tested individuals and $a\geq 0$ for untested individuals.)
	
	\subsection{The Growth Rate}
	The effects on the growth rate are also interesting. It is clear that $r = \frac{\sum_{j}a_j \dot{I}_j}{\sum_j a_j I_j}$ is only independent of $a$ if all $a_j$ are the same. Estimating the growth rate with imperfect information in an unstratified network model is for instance solved in \cite{UnderReporting6}.
	The growth rates for the individual groups $r_j = \frac{a_j \dot{I}_j}{a_j I_j} = \frac{\dot{I}_j}{I_j}$ can of course be estimated, but there is no simple relation between $r_j$ and $r$ because of the cross-terms between the groups. However, if a particular group $j'$ can be be said to be the clear driver of the outbreak, i.e.\ that their contribution to the dominant eigenvalue is the greatest, it is reasonable to assume $r \approx r_{j'}$, and therefore in this case possible to make an accurate estimate of the growth rate without knowing $a$. Note that $r_{j'}$ is not necessarily $\max_j r_j$, as a group with a very small number of infected can have a larger-than-natural growth rate if it primarily is driven by another group.
	
	\section{When is it possible to distinguish?}
	An obvious question is, when the above considerations break down. It turns out that if we add the possibility of reinfection, the transformation can no longer be done. That is, the model
	\begin{align}\label{Eq:SEIRS-model}
		\begin{split}
			\dot{S} &= - \diag(S)\beta I\\
			\dot{E} &=   \diag(S) \beta I + \diag(R) \diag(\mu) \beta I - \eta E\\
			\dot{I} &= \eta E - \gamma I\\
			\dot{H} &= \gamma p_H I - \sigma H\\
			\dot{R} &= (\eins-\diag(p_H))\gamma I + \sigma H - \diag(R) \diag(\mu) \beta I
		\end{split}\ .
	\end{align}
	Compared to Equation \eqref{Eq:SEIR-model}, we have added another vector with non-negative entries $\mu$, which accounts for reduced susceptibility of the previously infected.
	
	The transformation
	\begin{align}\label{Eq:SEIRS-model:Trans}
		\begin{split}
			S &= g \tilde{S}\\
			E &= g \tilde{E}\\
			I &= g \tilde{I}\\
			R &= g \tilde{R}\\
			p_H &= \tilde{p}_H g \inv\\
			\beta &= \tilde{\beta} g\inv
		\end{split}
	\end{align}
	keeps the dynamics approximately of all equations invariant for diagonal $g$ (where $\diag(S)$ and $g$ commute) and $p_H\ll 1$
	\begin{align}\label{Eq:SEIRS-model:Rescale}
		\begin{split}
			\dot{\tilde{S}} =& - \diag(\tilde{S}) \tilde{\beta} \tilde{I}\\
			\dot{\tilde{E}} =& \diag(S) \tilde{\beta} \tilde{I} + \diag(R) \diag(\mu) \tilde{\beta} \tilde{I} - \eta \tilde{E}\\
			\dot{\tilde{I}} =& \eta \tilde{E} - \gamma \tilde{I}\\
			\dot{H} =& \gamma \tilde{p}_H \tilde{I} - \sigma H\\
			\dot{\tilde{R}} =& (1 - g\inv\diag(\tilde{p}_H))\gamma I + g\inv\sigma H\\
			& - \diag(R) \diag(\mu) \beta I	
		\end{split}\ .
	\end{align}
	However, the normalization \eqref{Eq:Norm} becomes impossible. Alternative rescalings can also be done, but they cannot satisfy invariance of $S$ and $R$ and Equation \eqref{Eq:Norm} at the same time, even if we allow rescaling of $\mu$, unless the reinfection term $\diag(R) \diag(\mu) \tilde{\beta} \tilde{I}$ can be neglected in comparison to $\diag(S) \tilde{\beta} \tilde{I}$, such as in the initial phases of an outbreak.
	
	\section{Conclusion and Discussion}
	In the above we have shown that the traditionally available observables in humane epidemics are not sufficient to distinguish between a partially observed epidemic and a fully observed one with different parameters.
	
	Note the significance of these results. It means that unless the level of under reporting can be determined through other sources, it is impossible to estimate the attack rate in the population or the risk of hospitalization, even having observed the peak. The attack rate could be estimated through randomized screening test of the population to observe the current incidence, however in practice the turnout is far less than 100\%.
	In a recent study in Sweden the turnout was closer to 17\% \cite{Sweden}. There were also clear biases, e.g. a four times higher incidence among woman than men. Serological testing of blood donors \cite{SSIdonors} can provide an estimate of resent infections – even in a highly vaccinated population. However, that is neither a random sample of the population and does not cover all age groups.
	
	Furthermore, the under reporting will most likely vary in time due to availability of testing and guidelines. (E.g. there was a factor ~30 reduction in PCR tests in DK between falls of 2021 and 2022.) We have here assumed a constant level of under reporting. In reality, the level will change with time, but this only further complicates the matter.
	
	The amount of serious infections will of course still be known through the hospital admissions, but if one wants to estimate the risk of long-term effects in the population, such as long covid and the immunity towards new variants, the disease dynamics are never sufficient.

	\acknowledgements
	\paragraph{Acknowledgements.} This work was funded by Statens Serum Institut, Denmark. (AM)

\end{document}